# An Improved Approach to High Level Privacy Preserving Itemset Mining

Rajesh Kumar Boora         Ruchi Shukla[$]         A. K. Misra

Computer Science and Engineering Department
Motilal Nehru National Institute of Technology, Allahabad, India – 211004

([$] Corresponding author)

*Abstract*—Privacy preserving association rule mining has triggered the development of many privacy-preserving data mining techniques. A large fraction of them use randomized data distortion techniques to mask the data for preserving. This paper proposes a new transaction randomization method which is a combination of the fake transaction randomization method and a new per-transaction randomization method. This method distorts the items within each transaction and ensures a higher level of data privacy in comparison to the previous approaches. The per-transaction randomization method involves a randomization function to replace the item by a random number guarantying privacy within the transaction also. A tool has also been developed to implement the proposed approach to mine frequent itemsets and association rules from the data guaranteeing the anti-monotonic property.

*Keywords; Data Mining, Privacy, Randomization, Association Rules*.

## I. INTRODUCTION

Data mining deals with the problem of discovering unknown patterns from data. It includes building models on data, presenting statistical summary of data in human understandable form, deciding upon strategies based on the extracted information. The main consideration in privacy preserving data mining is the sensitive nature of raw data. The data miner, while mining for aggregate statistical information about the data, should not be able to access data in its original form with all the sensitive information. This call for more sophisticated techniques in privacy preserving data mining that intentionally modify data to hide sensitive information, but still preserve the inherent statistics of the data important for mining purpose. Randomization is the only effective approach to preserve the privacy in a system with one data miner and multiple data providers. The discovery of interesting association relationships among huge amounts of business transaction records can help catalog design, cross-marketing, loss-leader analysis, and other business decision making processes.

The main goal of the association rule mining is to find out associations or correlations between the items of the particular data involved in the mining. An association rule is an implication of the form $X => Y$ where $X, Y \subset I$ are sets of items called itemsets and $X \cap Y = \Phi$. Association rule mining finds the frequent itemsets of a data based on two measurements: *Support* and *Confidence*.

*Definition 1:* Let $I$ be a set of n items: $I = \{a_1, a_2, \ldots, a_n\}$. Let $T$ be a sequence of $N$ transactions: $T = \{t_1, t_2, \ldots, t_n\}$ where each transaction $t_i$ is a subset of $I$. Given an itemset $A \subseteq I$, the *support* of $A$ is defined as

$$\mathrm{supp}^T(A) = \frac{\#\{t \in T \mid A \subseteq T\}}{N}. \quad (1)$$

If $\mathrm{supp}^T(A) \geq S_{\min}$, then A is a frequent itemset in $T$, where $S_{\min}$ is a user-defined parameter called minimum support [3].

*Definition 2:* The *confidence* for an association rule $X => Y$ is the ratio of the number of transactions that contain $X \cup Y$ to the number of transactions that contain $X$.

$$\textit{Confidence of an association rule } X => Y = \frac{\#\{X \cup Y\}}{\#\{X\}}. \quad (2)$$

In this work, a new approach to privacy preserving data mining has been proposed. This approach is a mixture of the fake transaction randomization method and a new proposed per-transaction randomization method. The fake transaction randomization method adds fake transactions randomly in between the real transactions [2]. This approach provides good mining results with small probability of error and guaranteed data privacy. However, in recent years, most of the research has focused on privacy preserving data mining using the per-transaction randomization method. Hereby, a different per-transaction randomization approach is proposed, which includes a randomization function to distort each item of every transaction and does not influence the support of any itemset. Besides, reconstruction function applied on the distorted data produces near absolute accurate items. A tool has also been developed to implement the new approach.

## II. RELATED WORK

Many recent publications on privacy have focused on the perturbation model. A perturbation based approach was proposed by Agrawal and Srikant which built a decision-tree classifier from training data [1]. References [4] and [6] presented the problem of association rule mining, where transactions are distributed across multiple sites. References [1] and [7] mine the perturbed data instead of original data and [8] used cryptographic techniques to preserve the privacy in distributed scenarios. Reference [9] presented that the randomization approach can be used to determine web demographics, while the cryptographic approach can be used to support inter enterprise data mining [2]. Reference [10] proposed an algebraic technique to preserve the privacy, which results in identifying association rules accurately but discloses less private information. A growing body of literature exists on different approaches of privacy preserving data mining. Some of these approaches adopted for privacy preserving data mining are briefly summarized below:

### A. P3P and Secure Database

P3P covers the system and architecture design perspectives of privacy preserving data mining. It does not involve any development of algorithm for data mining on sensitive data. P3P provides a way for web site owners to encode their privacy policies in a standard XML format so that users can check against their privacy preferences to decide whether or not to release their personal data to the web site. Koike [11] provides a detailed survey of current P3P implementations. The basic P3P architecture is client based, i.e. privacy of client is defined at web-client end. As opposed to the client-centric implementations, the author in [12] proposed a server-centric architecture for P3P. The work presented in [13] addressed the problem of enforcing the web sites act according to their stated privacy policies by identifying the technical challenges and founding principles in designing a Hippocratic database.

### B. Secure Multi-Party Computation

Secure Multi-Party Computation (SMC) is the problem of evaluating a function of two or more parties' secret inputs. Each party finally holds a share of the function output. No extra information specific to a party is revealed except what is implied by the party's own inputs and outputs [14]. The work described in [16] proposed a paradigm of information sharing across private databases based on cryptographic protocols. Compared with the brute force circuit schema, this algorithm is much faster. The work in [17] described several secure multi-party computation based algorithms that can support privacy preserving data mining, e.g., secure sum, secure set union, secure size of set intersection and secure scalar product. The computation is secure if the view of each site during the execution of the protocol can be effectively simulated by the input and the output of the site. This is not the same as saying that private information is protected [6].

### C. Data Swapping

Data swapping is a simple technique to preserve confidentiality of individual values of sensitive data without changing the aggregate statistical information of the data. The basic idea is to transform the database by switching a subset of attributes between selected pairs of records. As a result, the lower order frequency counts or marginals are preserved and data confidentiality is not compromised [15].

### D. Privacy Preserving Distributed Data Mining

Kantarcioglu and Clifton proposed the privacy preserving distributed mining of association rules on horizontally partitioned data [6]. They considered the individual sites: $S_i = \{S_1, S_2, \ldots, S_n\}$. The criterion is that the each site calculates the locally frequent itemsets and these results are securely transmitted to the global site. Then the global site calculates the globally frequent itemsets. Each and every site calculates the support of itemsets. An algorithm for privacy preserving mining of association rules in distributed databases that builds a global hashing table $H_i$ in every iteration, is proposed by Liu [18].

### E. Data Distortion

The algorithms belonging to this group work by first distorting the data using randomized techniques. The perturbed data is then used to extract the patterns and models for reconstructing the support of items from perturbed data. The perturbation approach results in some amount of information loss but larger perturbations also lead to a greater amount of privacy. Thus there is a natural trade-off between greater accuracy and loss of privacy [1]. Distortion using multiplicative noise to protect confidentiality of the data is also considered as another option in data mining. The first approach is based on generating random noise that follows truncated normal distribution with unit mean and small variance, and multiplying each element of the original data by the noise. The second approach is inspired by additive random perturbation in a logarithmic domain [7]. The data undergoes a logarithmic transformation, and random noise is generated following a multivariate normal distribution with mean zero and constant variance [19]. The non-negative matrix factorization (NMF) with sparseness constraints for data perturbation is proposed to provide the privacy [20].

## III. FAKE TRANSACTION RANDOMIZATION METHOD

The fake transaction randomization method generates fake transactions randomly in between the real transactions based on the two characteristics: quantity and quality of fake transactions. The goal of the fake transaction method is to preserve the privacy of critical customer data. As long as these fake transactions look like real transactions and the number of fake transactions is similar to or exceeds the number of real transactions, the privacy of the real transactions is preserved. If the distribution of the lengths of the real transactions differs from that of the fake transactions, privacy breaches are likely. Consequently, if the distribution of the lengths of the real transactions is known and used as a parameter for generating fake transactions, then the quality of the fake transactions can be markedly improved [2].

Given *w* and the number of the real transactions '*N*' the number of fake transactions is *wN*. *X* fake transactions are inserted between every two real transactions, where *X* is a random variable that is uniformly distributed with a mean of *w*, a minimum of 0, and a maximum of 2*w*. The steps by which a fake transaction is generated are as follows. First, the length of the fake transaction is determined using a uniformly distributed random variable *Y*, whose mean, minimum, and maximum are *l*, 1 and 2*l* -1, respectively. Then, *Y* distinct items are randomly selected from *I* and these *Y* items are finally used to generate a fake transaction [2]. If both the mean and the variance of the lengths of the real transactions are known, then the length of the fake transactions can be allowed to follow a normal distribution with the same mean and variance, to improve further the fake transactions.

### A. Example 1

Suppose the items provided by the supermarket are *green apples, red apples, oranges, bananas* and *grapes*. Consider it as the set of items *I* = {*green apples, red apples, oranges, bananas, grapes*}. Transactions of the super market are given in Table I.

TABLE I. TRANSACTIONS OF SUPERMARKET DATA WITH INTEGER NUMBER REPRESENTATION

| Transaction | Customer | Items bought (transactions) | Positive number representation |
|---|---|---|---|
| T1 → | C1 | green apples, grapes | 1, 5 |
| T2 → | C2 | oranges | 3 |
| T3 → | C3 | oranges, grapes | 3, 5 |
| T4 → | C4 | red apples, bananas | 2, 4 |
| T5 → | C5 | bananas | 4 |
| T6 → | C6 | green apples, red apples | 1, 2 |
| T7 → | C7 | green apples, oranges | 1, 3 |
| T8 → | C8 | oranges, green apples, grapes | 3, 1, 5 |

The new proposed method represents the items in the super market with the positive numbers (1- green apples, 2 - red apples, 3 - oranges, 4 – bananas and 5 – grapes). The super market transactions are replaced by the corresponding positive numbers. In the mining process, the transactions are having the integer numbers as well as strings. It is somewhat difficult to deal with different data types, so it is better to consider all items of transactions with the positive numbers. This makes the mining process easier to produce the results.

Now, the fake transactions have to be added to the above real transactions to preserve the privacy. To generate the fake transactions some major characteristics have to be considered. The average length of the fake transactions is determined by calculating the average length of the real transactions and assigns it by an almost near value equal to the average length of the real transactions. To calculate the average length of the real transactions, first find out the length of the each real transaction, which is equal to the number of items involved in that particular transaction. Average length of the real transaction = sum of lengths of real transactions / total number of transactions.

In Example 1, the average length of fake transaction (*l*) is 2. Let *w* is the ratio of number of fake transactions to the number of real transactions. Consider that three fake transactions will be added for every two real transactions. The total number of fake transactions (*wN*) =12, where *N* is the total number of real transactions. Therefore, 12 fake transactions are generated and mixed with the *N* real transactions. Table II shows the mixed transactions in the Example 1.

TABLE II. SUPERMARKET DATA WITH MIXED TRANSACTIONS

| Transaction | Items bought (transactions) |
|---|---|
| T1 → | 1, 5 |
| T2 → | 3 |
| T3 → | 1, 4 |
| T4 → | 2 |
| T5 → | 5, 4 |
| T6 → | 3, 5 |
| T7 → | 2, 4 |
| T8 → | 1 |
| T9 → | 3, 1 |
| T10 → | 5, 2 |
| T11 → | 4 |
| T12 → | 1, 2 |
| T13 → | 3 |
| T14 → | 4, 2 |
| T15 → | 1, 3 |
| T16 → | 3, 1, 5 |
| T17 → | 5, 3 |
| T18 → | 3 |
| T19 → | 4, 5 |
| T20 → | 1, 4 |

So, the probability of selecting a real transaction from the mixed transactions is very less. Therefore, the privacy is guaranteed in between the transactions by mixing the fake transactions to the real transactions.

### B. Limitations of Fake Transaction Randomization Method

The limitation of fake transaction randomization method is that it achieves the privacy up to the level of transaction to transaction only but not within the transactions. That means, this method adds the fake transactions in between the real transactions and it would not distort the any of the items involved in the transactions. By distorting the items within the transactions we can achieve an even higher level of privacy than the earlier. So, we proposed a new per-transaction randomization method to achieve higher privacy.

## IV. PROPOSED PER-TRANSACTION RANDOMIZATION METHOD

This method involves the modification of data items and can be applied on both the real and fake transactions. It adds

(IJCSIS) International Journal of Computer Science and Information Security,
Vol. 6, No. 3, 2009noise directly to the data items involved within the transactions. Specifically, a per-transaction randomization function is proposed which modifies the data items within the transaction.

A. *Per-Transaction Randomization Function:*

Let *R* is the integer number generated by per-transaction randomization function. The data item within the transaction is replaced by *R*.

$$R = (item + tnoi + i) \% tnoi. \qquad (3)$$

Where, *item* is the data item within the transaction, *tnoi* is the total number of items i.e. |*I*|, and *i* is the random number generated from the random number generator and it is fixed during the entire randomization process. The per-transaction randomization function is applied on each and every data item within the transaction to modify each data item. The process of per-transaction randomization is applied after the fake transaction method, so that a real data item within the transaction can not be identified. The main advantage of this function is that it does not affect the support of any itemset in the mixed transactions. By applying these two methods, privacy is guaranteed between the transactions as well as within the transaction.

Suppose, the per-transaction method is applied before the fake transaction randomization method, then there is a large variation occurred between the real and the fake transactions. If the per-transaction randomization method is applied first, then it modifies the real transactions by adding some noise to the data items involved in the transaction. After that, fake transaction randomization method applied, and then the fake transactions are mixed with the resultant transactions of the per-transaction randomization method. The problem here is that, actually the fake transaction randomization method uses *I* (set of items) to generate the fake transactions. That is, in the mixed transactions only some transactions are took the affect of per-transaction randomization method. In the mining process, at the time of finding out the support of any item, it is resulting in support value error. That is the reason the fake transaction randomization method is applied before the per-transaction randomization method. Therefore, a high level privacy of the customer data is achieved.

B. *Example 1*

After the fake transaction randomization method is applied on Example 1 which is shown in section 3, the per-transaction randomization method is applied on the resultant (Example 1) of fake transaction randomization method. In Example 1, total number of items (*toi*) = 5, and considered that *i* = 4, which is generated by the random number generator. The per-transaction randomization function is applied on each transaction in Table II and the resultant transactions are given in Table III. Thus after applying the present approach a higher-level of privacy is provided to the customer data.

TABLE III. SUPERMARKET DATA AFTER PER-TRANSACTION RANDOMIZATION METHOD

| Transaction | Items bought (transactions) |
|---|---|
| T1 → | 5, 4 |
| T2 → | 2 |
| T3 → | 5, 3 |
| T4 → | 1 |
| T5 → | 4, 3 |
| T6 → | 2, 4 |
| T7 → | 1, 3 |
| T8 → | 5 |
| T9 → | 2, 5 |
| T10 → | 4, 1 |
| T11 → | 3 |
| T12 → | 5, 1 |
| T13 → | 2 |
| T14 → | 3, 1 |
| T15 → | 5, 2 |
| T16 → | 2, 5, 4 |
| T17 → | 4, 2 |
| T18 → | 2 |
| T19 → | 3, 4 |
| T20 → | 5, 3 |

V. SUPPORT RECONSTRUCTION

During the reconstruction of itemsets after fake transaction randomization method and per-transaction randomization method, the items in the transactions are modified to preserve the privacy. Hence, the support of frequent itemsets are affected (i.e., support values are modified). So, there is a need to reconstruct the support of frequent itemsets from the mixed transactions. In order to achieve 100% support of itemsets in data the fake transaction and per-transaction reconstruction methods are applied and are discussed in the following sections.

A. *Support Reconstruction of Fake Transaction Method*

At this point, the transactions of any data are the resultant of both the fake transaction randomization method and per-transaction randomization method. In this section, reconstruction procedure of fake transaction randomization method is discussed. For a given k-itemset *A*, the number of real transactions that support *A* in mixed transactions is given as follows [2]:

The number of real transactions that support *A* = the number of mixed transactions that support *A* – the number of fake transactions that support *A*. Let *S`* is the support of some k-itemset *A* in mixed transactions *T`*, that is $supp^T(A) = S`$. Then, the number of transactions that support *A* is *S`*(1+*w*)*N*, where (1+*w*)*N* is the number of transactions in *T`*. Let support of itemset *A* in *T* by *S*, i.e., $supp^T(A) = S$. Therefore, *S* can be derived from *S`* as follows: (excerpted from [2])

$$S = S`(1+w) - \frac{w}{nCk(2l-1)} \sum_{Y=k}^{2l-1} YCk \qquad (4)$$

219

### B. Reconstruction of the Per-transaction Randomization Method

After reconstructing the support from the fake transaction method, the next problem is to reconstruct the original items from per-transaction randomization method. So, a per-transaction reconstruction function is proposed to solve the above problem. The reconstruction function is applied on the result of fake transaction reconstruction method i.e., the per-transaction reconstruction function is applied on frequent itemsets. The per-transaction reconstruction function is presented as follows:

$$Oitem = (R - i + |I|) \% |I| \quad (5)$$

where, *Oitem* is the original item after the reconstruction function, *R* is the frequent item on which per-transaction randomization function is applied, *i* is the random number which is passed from random number generator and it is fixed entire the process of reconstruction method and |*I*| is the total number of items.

The per-transaction randomization function does not affect the support value of any item in the transactions and it just modifies the original item by adding the noise. The per-transaction reconstruction function reconstructs the original items without any probability of error i.e., the percentage of error for per-transaction reconstruction method is zero. There is no need to apply the per-transaction reconstruction function on the every item in the mixed transactions. That is the per-transaction reconstruction function is applied only on the result of the mining process i.e., frequent items. Therefore, the per-transaction randomization method provides higher accuracy in finding frequent itemsets and guarantees a higher level of privacy to the customer data.

### C. Example

To find the frequent itemsets in Example 1, the data mining process is applied on transactions in Table II which is real transaction database and Table III (mixed transactions) which is the resultant of fake and per-transaction randomization methods. The data mining process is applied to each item in the set *I* (set of items). As per definition 1, Support of an item =

$$\frac{\text{the number of occurrences of item in transactions}}{\text{total number of transactions}}$$

Considering the Example 1, it has 5 items in the item set *I*. Let $S_{min} = 0.4$, which is the minimum support provided by the user. Suppose if any item is having the support value greater than or equal to the minimum support, then that item is frequent item. First, the data mining process is applied on Table II i.e. real transactions. The supports for 5 items in real transactions are calculated. Support of items in real transactions $S_1 = 0.5$, $S_2 = 0.25$, $S_3 = 0.5$, $S_4 = 0.25$ and $S_5 = 0.385$. The support for item 1 ($S_1$) = 0.5 ≥ 0.4 ($S_{min}$) and support for item 3 ($S_3$) = 0.5 ≥ 0.4 ($S_{min}$). Therefore, items 1 and 3 are frequent 1-itemsets in real transaction database.

To find out the support of items of real transactions in mixed transactions, first the mining process has to find out the support ($S'$) of items in mixed transactions. Now, the data mining process is applied on Table III that is mixed transactions. The supports for 5 items in mixed transactions are calculated.

Support of items in mixed transactions $S'_1=0.25$, $S'_2=0.4$, $S'_3=0.35$, $S'_4=0.35$ and $S'_5=0.4$. The support for item 2 ($S'_2$) = 0.5 ≥ 0.4 ($S_{min}$) and support for item 5 ($S'_5$) = 0.5 ≥ 0.4 ($S_{min}$). Therefore, items 2 and 5 are frequent 1-itemsets in mixed transaction database. Now, it has to be check that whether the items 2 and 5 are frequent items of real transactions in mixed transactions or not. It is considered that, $n = 5$, which is number of items (|*I*|), $N = 20$, which is number of real transactions, $k = 1$, that is k-itemset is having only one item, $l = 2$, average length of fake transactions, $w = 3/2$, ratio of number of fake transactions to the number of real transactions, and $S' = 0.4$ for both the items 2 and 5, using the Eq. 3. For item 2, $S = 0.4 \geq S_{min}$, and for item 5, $S = 0.4 \geq S_{min}$. For both of the items, the support values are greater than equal to the user-defined minimum support ($S_{min}$). As per the Definition 1, both the items are frequent items of real transactions in mixed transactions. Earlier, the per-transaction randomization method is applied on mixed transactions, so the frequent items 2 and 5 are modified items. To get the reconstructed items, the per-transaction reconstruction method has to be applied on these two frequent items.

Eq. 3 is used to find out the reconstructed frequent itemsets. In Example 1, it is considered that total number of items |*I*| = 5 and $i = 4$ which is passed to the per-transaction reconstruction function. The per-transaction reconstruction function is applied on items 2 and 4, and it results that, items 1 and 3 are original frequent items of real transactions in mixed transactions. In real transaction database also the items 1 and 3 are frequent items. So, it is concluded that both the reconstructed methods produced the same items (1 and 3) as the frequent items. So, in Example 1, the items 1 and 3 correspond to the *green apples* and *oranges*. Therefore, the items *green apples* and *oranges* are frequent items.

### VI. TOOL AND EXPERIMENTAL RESULTS

In this section, the snapshots of the tool developed are presented to find out the frequent itemsets and association rules from the transactions of any dataset. The snapshots are taken when the mining process is working with the CSC dataset. The CSC dataset results (i.e., frequent itemsets) are also shown in the snapshots of the tool.

Figure 1 is the snapshot of the menu screen in the tool. It allows the user to select any particular functionality, like generating fake transactions, applying per-transaction randomization method, generating database, finding frequent itemsets and finding association rules.

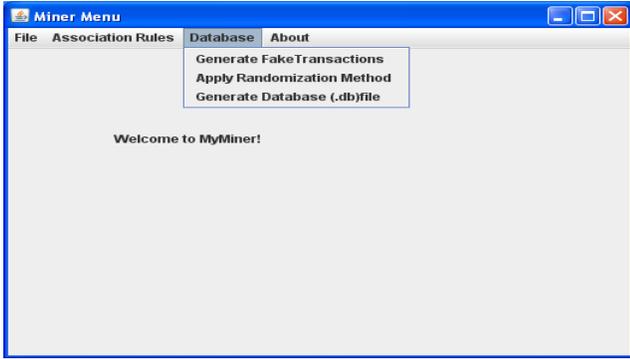

Figure 1. Snapshot of the Menu screen of the tool

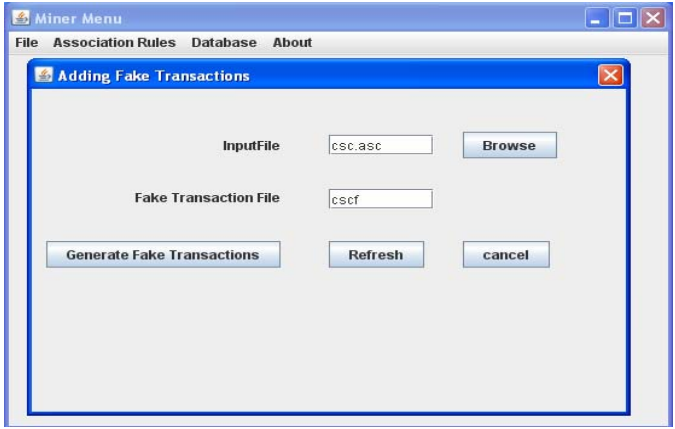

Figure 2. Generating fake transaction file

Figure 2 shows how to generate the fake transaction file. The snapshot allows the user to select the real transaction file through the 'Browse' button and also allows the user to mention the fake transaction file name that has to be generated. The term 'user' corresponds to a company's authorized person, who is appointed by the company to use the tool. On clicking the 'Generate Fake Transactions' button, it adds the fake transactions with the real transactions and privacy is provided in between the transactions. The snapshot for applying per-transaction randomization approach is also same as the above snapshot shown in Figure 2.

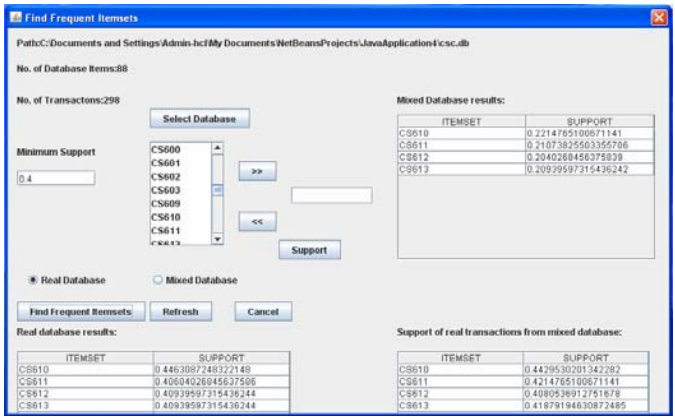

Figure 3. Finding frequent itemsets in the database

Figure 3 is the screen shot of the tool to find the frequent itemsets in the database and this screen shot using three tables to compare the results. Table III shows the frequent itemset and corresponding support of real transactions in mixed transaction database and Table IV shows the support of frequent itemsets in the real transaction database. The user must select the appropriate radio button when working with the different databases (like real transactions database and mixed transactions database).

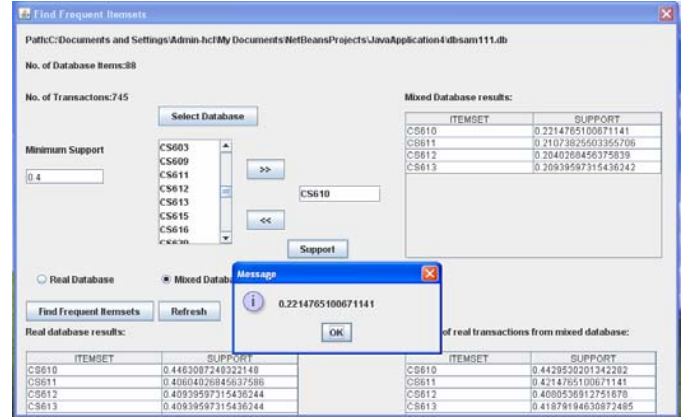

Figure 4. Finding support of individual itemset

Figure 4 shows the support of a particular itemset given by user i.e. it allows the user to find the support of any particular itemset. Figure 5 shows the screen shot of the tool to find the association rules. It allows the user to provide the minimum support, and minimum confidence and it produces the association rules with corresponding support and confidence values.

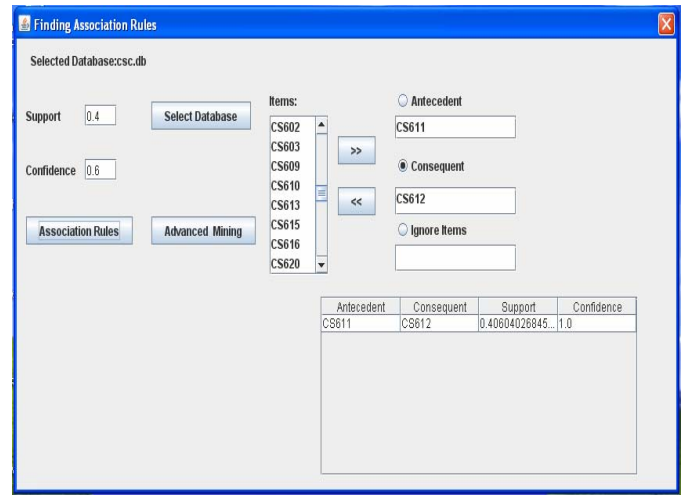

Figure 5. Finding Association Rules

## VII. EXPERIMENTAL RESULTS

To test and validate the developed tool, experiments were conducted on the CSC and mushroom datasets. The CSC

dataset have nearly 300 real transactions, 88 data items. The average length of fake transactions ($l$) is set to the integer which is closer to the average length of real transactions. It is considered that $w=2$ and $S_{min}=0.4$, then the mining process is applied to find out the support of frequent itemsets. The support of frequent itemsets in real transaction database and support of frequent itemsets of real transactions in mixed transaction database is obtained and is shown in Table IV.

TABLE IV. SUPPORT OF FREQUENT ITEMSETS IN CSC DATASET

| Itemset | Support in Real Transactions Database | Support of Real transactions in Mixed Transactions Database |
|---|---|---|
| CS610 | 0.4463087248322148 | 0.4429530201342282 |
| CS611 | 0.40604026845637586 | 0.4214765100671141 |
| CS612 | 0.40939597315436244 | 0.4080536912751678 |
| CS613 | 0.40857629514354242 | 0.41879194630872485 |

Consider the frequent item CS611, it is having a small support difference of 0.015436 between the real and mixed transaction database. This difference would not make any problem while finding the frequent itemsets because the support values in the real and mixed transactions is greater than the user-defined minimum support (i.e., $S_{min} = 0.4$) and is negligible.

Figure 6 shows the closeness of supports of different itemsets in real and mixed transactions database (of CSC dataset). It can be concluded that the support values of frequent itemsets are almost equal in real and mixed transaction databases leading to the inference that the probability of error is infinitesimally small and can be treated as negligible. Therefore, accurate mining results are obtained, highlighting the novelty of the proposed approach.

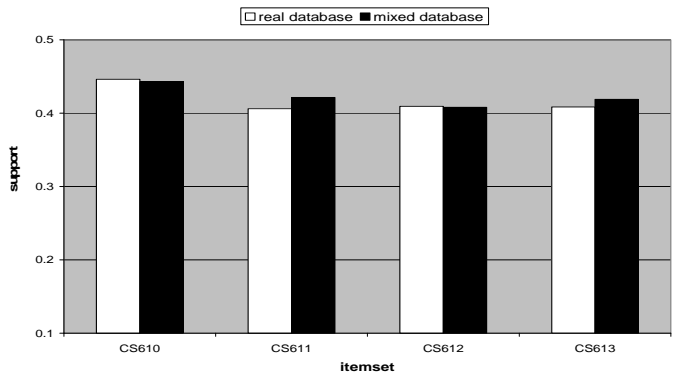

Figure 6. Comparison of supports of items in CSC dataset

Figure 7 shows the comparison between the new approach and the approach of Lin and Liu. The arrows $a1$, $a2$, $a3$ and $a4$ in Figure 7 indicate the faults in reconstructing the items using the previous approach. At $a1$, the original item is 10 but the reconstructed item is 7, at $a2$, the original item is 12 but the reconstructed item is 14, and so on, using the pervious method. The per-transaction randomization approach produces accurate reconstructed items for corresponding original items (i.e., 10 for 10, 19 for 19 …). Therefore, the proposed per-transaction randomization approach guarantees privacy within the transactions with improved accuracy.

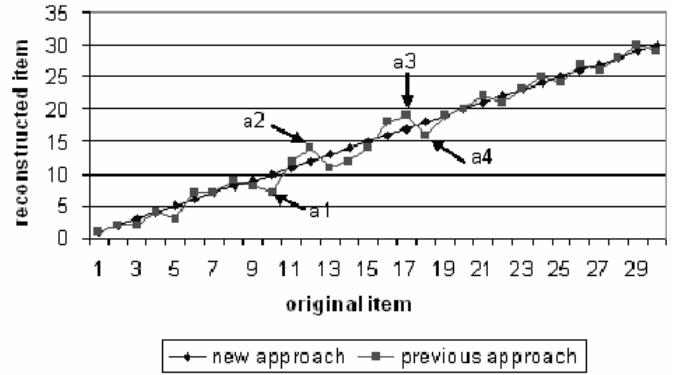

Figure 7. Comparison of previous and new per-transaction randomization methods

The mining process is also applied on mushroom dataset of 8124 real transactions and 28 data items. The average length of fake transactions ($l$) is set to the integer which is closer to the average length of real transactions. It is considered that $w = 2$ (i.e., 3 fake transactions are added for every 2 real transactions) and $S_{min} = 0.9$, then the mining process is applied to find out the support of frequent itemsets. The support of frequent itemsets in real transaction database and real transactions in mixed transaction database is obtained as shown in Table V.

TABLE V. SUPPORT OF FREQUENT ITEMSETS IN MUSHROOM DATASET

| Itemset number | Itemset | Support in Real Transactions Database | Support of Real transactions in Mixed Transactions Database |
|---|---|---|---|
| 1 | gill-attachment=free | 0.9741506646480 | 0.96878388503200 |
| 2 | veil-type=partial | 1.0 | 0.9937961595273 |
| 3 | veil-color=white | 0.97538158542 | 0.9730182176267 |
| 4 | ring-number=one | 0.92171344165 | 0.9293943870014 |
| 5 | gill-attachment=free, veil-type=partial | 0.97415066469 | 0.918513047759 |
| 6 | gill-attachment=free, veil-color=white | 0.973165928114 | 0.917134416543 |
| 7 | veil-type=partial, veil-color=white | 0.975381585425 | 0.9218611521418 |
| 8 | gill-spacing=close | 0.8385032003938 | 0.858493353028065 |

Table V shows the closeness of supports of different itemsets in real and mixed transactions database (of mushroom dataset). It can be concluded that the support values of frequent itemsets are almost equal in real and mixed transaction databases leading to the inference that the probability of error is infinitesimally small and can be treated as negligible.

## VIII. CONCLUSIONS AND FUTURE WORK

In this work, a new approach for privacy preserving association rule mining is presented. This approach provides excellent accuracy in reconstructing frequent itemsets with no influence on support of itemsets. At present, the application of this approach is limited to a local environment i.e., used within the organization. This work can be extended further to deal with a distributed environment.

AUTHORS PROFILE

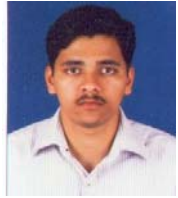

Rajesh Kumar Boora is an MTech in Software Engineering from the Motilal Nehru National Institute of Technology, Allahabad, India and is currently serving as an Associate Application Developer at CSC, Hyderabad, India. His research interests include data mining and artificial intelligence. He has around 2 years of experience in industry and has published papers in international conferences.

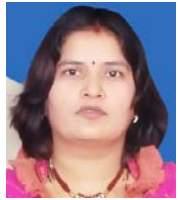

Ruchi Shukla is an MTech in Software Engineering and is currently pursuing Ph.D. in Computer Science and Engineering from the Motilal Nehru National Institute of Technology, Allahabad, India. Her research interests include software engineering, data mining, effort estimation and artificial intelligence. She has over 5 years of experience in teaching and research and has published papers in international conferences and journals.

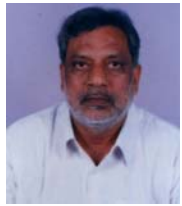

A K Misra is a Ph.D. in Computer Science and Engineering from the Motilal Nehru Regional Engineering College, Allahabad, India and is currently the Professor and Head of the Department of Computer Science and Engineering at Motilal Nehru National Institute of Technology, Allahabad, India. His research interests include software engineering, artificial intelligence and data mining. He has over 35 years of experience in teaching, research and administration and has supervised many graduate and doctoral students. To his credit he has over 65 papers in international journals and conferences.